\documentclass[prd,aps,floats,twocolumn,showpacs,nofootinbib,superscriptaddress,preprintnumbers,showkeys]{revtex4}

\usepackage[latin1]{inputenc}                    
\usepackage{graphicx}                            
\usepackage{latexsym}                            
\usepackage{amsfonts}                            
\usepackage{amssymb}                             
\usepackage{amsmath}                             
\usepackage[mathscr]{eucal}                      
\usepackage{dcolumn}                             
\usepackage{theorem}                             
\usepackage{hyperref}                            
\graphicspath{{.}{Figs/}}                    
\newcommand{\msbar}{$\overline{\mbox{\rm MS}}\ $}  
\newcommand{\cO}{\mathcal {O}}
\newcommand{\Dd}[1]{\overset{\leftrightarrow}{D}^{#1}}
\newcommand{\be}{\begin{equation}}
\newcommand{\ee}{\end{equation}}

\newcommand{\del}{\partial}
\newcommand{\gev}{{\,{\rm GeV}}}
\newcommand{\xm}{\langle x^m\rangle}
\newcommand{\x}{\langle x\rangle}
\newcommand{\1}{\langle 1\rangle}
\newcommand{\mqbar}{\overline{m}_q}
\newcommand{\nc}{\newcommand}
\nc{\CSV}{{\scriptscriptstyle CSV}}
\nc{\EA}{{{\it et al.}}}

\begin{document}

\preprint{
\vbox{
\hbox{ADP-12-11/T778,  DESY 12-052, Edinburgh 2012/02, LTH 942}
}}

\title{Charge Symmetry Breaking in Spin Dependent Parton Distributions
  and the Bjorken Sum Rule}

\author{I.~C.~Clo{\"e}t}
\affiliation{CSSM, School of Chemistry and Physics, University of Adelaide, Adelaide SA 5005, Australia}

\author{R.~Horsley}
\affiliation{School of Physics and Astronomy, University of Edinburgh,
  Edinburgh EH9 3JZ, UK}

\author{J.~T.~Londergan}
\affiliation{Department of Physics and Center for Exploration of
             Energy and Matter, Indiana University, Bloomington, IN 47405, USA}

\author{Y.~Nakamura}
\affiliation{RIKEN Advanced Institute for Computational Science, Kobe, Hyogo 650-0047, Japan}

\author{D.~Pleiter}
\affiliation{J\"ulich Research Centre, JSC, Room 324, 52425 J\"ulich,Germany}
\affiliation{Institut f\"ur Theoretische Physik, Universit\"at Regensburg, 93040 Regensburg, Germany}

\author{P.~E.~L.~Rakow}
\affiliation{Theoretical Physics Division, Department of Mathematical
             Sciences, University of Liverpool, Liverpool L69 3BX, UK}

\author{G.~Schierholz}
\affiliation{Deutsches Elektronen-Synchrotron DESY, 22603 Hamburg, Germany}

\author{H.~St\"uben}
\affiliation{Konrad-Zuse-Zentrum f\"ur Informationstechnik Berlin, 14195 Berlin, Germany}

\author{A.~W.~Thomas}
\affiliation{CSSM, School of Chemistry and Physics, University of Adelaide, Adelaide SA 5005, Australia}
\affiliation{Centre of Excellence for Particle Physics at the
  Terascale, School of Chemistry and Physics, University of Adelaide,
  Adelaide SA 5005, Australia}

\author{F.~Winter}
\affiliation{School of Physics and Astronomy, University of Edinburgh,
  Edinburgh EH9 3JZ, UK}

\author{R.~D.~Young}
\affiliation{CSSM, School of Chemistry and Physics, University of Adelaide, Adelaide SA 5005, Australia}
\affiliation{Centre of Excellence for Particle Physics at the
  Terascale, School of Chemistry and Physics, University of Adelaide,
  Adelaide SA 5005, Australia}

\author{J.~M.~Zanotti}
\affiliation{CSSM, School of Chemistry and Physics, University of Adelaide, Adelaide SA 5005, Australia}

\collaboration{CSSM and QCDSF/UKQCD Collaborations}

\begin{abstract}
  We present the first determination of charge symmetry violation
  (CSV) in the spin-dependent parton distribution functions of the
  nucleon.
  This is done by determining the first two Mellin moments of the
  spin-dependent parton distribution functions of the octet baryons
  from $N_f = 2 + 1$ lattice simulations.
  The results are compared with predictions from quark models of
  nucleon structure.
  We discuss the contribution of partonic spin CSV
  to the Bjorken sum rule, which is important because the
  CSV contributions represent the only partonic corrections to the
  Bjorken sum rule.
\end{abstract}


\maketitle

\section{Introduction}
Charge symmetry refers to the invariance of the strong interaction
under a very particular operation in isospin space, namely the interchange of $u$ and $d$ quarks
and also protons and neutrons.
Technically, the charge symmetry operator $P_{CS}$ corresponds to a
rotation of $180^{\circ}$ about the $2$ axis in isospin space.
In nuclear systems, charge symmetry is generally valid to 
substantially better than 1\% \cite{Miller:2006tv}. 
At the partonic level, charge symmetry implies the equality of
different parton distribution functions (PDFs), namely
\be u^p(x,Q^2) = d^n(x,Q^2), \hspace{0.6cm} d^p(x,Q^2) = u^n(x,Q^2),
\label{eq:csvPDF}
\ee
with analogous relations for antiquark PDFs.
To date, no experimental violation of charge symmetry has been
observed at the partonic level, and current upper limits are
consistent with the validity of partonic charge symmetry in the range
5-10\% \cite{Londergan:1998ai,Londergan:2009kj}. 

In this letter we report the first determination of the charge 
symmetry violation (CSV) in the
spin-dependent parton distributions arising from quark mass
differences. 
We begin by extracting the zeroth and first moments of the
spin-dependent PDFs of the light baryon octet by varying the light
(degenerate $u,d$) and strange quark masses in a $N_f=2+1$ lattice
simulation. We compare these results to quark model predictions for
the sign and magnitude of these moments.
Finally, we examine the size of the expected contribution of the spin
parton distributions to the Bjorken Sum Rule.

\section{Charge Symmetry Violation}
Theoretical models for partonic charge symmetry predict that the
spin-independent parton CSV distributions
$\delta u^-(x) = u^{p-}(x) - d^{n-}(x)$ and $\delta d^-(x) = d^{p-}(x)
- u^{n-}(x)$ should be roughly equal in magnitude and opposite in sign
\cite{Londergan:2009kj}, where the minus superscript denotes the
valence or C-odd combination of parton distribution functions,
\be
q^\pm(x) = q(x) \pm \bar{q}(x)\ .
\ee

The MRST group included valence CSV in a
global analysis of high energy experimental data
\cite{Martin:2003sk}. 
The best value obtained in this search was in excellent agreement with
quark model calculations of valence parton CSV \cite{Rodionov:1994cg},
but with very large errors.
A recent lattice calculation was able to probe the magnitude of
CSV violation \cite{Horsley:2010th}. 
There, the behaviour of the first moments of the hyperon parton
distribution functions were studied as the light and strange quark
masses were varied in a $N_f = 2 + 1$ lattice simulation.
The first moment of the parton distributions $\delta u^+(x)$ and
$\delta d^+(x)$ agreed very well in sign and magnitude with both the
quark model results and the best value from the global fit --- with
the uncertainties on the lattice results substantially smaller than
those from the global fit.
Note that the lattice calculation accessed the C-even combinations of
partonic CSV distributions, so the lattice results contained some sea
quark CSV effects that were not included in the other investigations.

We define the $m^\mathrm{th}$ moment of the charge symmetry violating
spin-dependent quark distributions in the nucleon as
\begin{align}
\label{eq:delu}
\delta \Delta u^{m} &= \int_0^1 dx\, x^m (\Delta u^{p}(x)- \Delta d^{n}(x)), \nonumber \\
                   &= \xm^p_{\Delta u} - \xm^n_{\Delta d}\,, \\
\label{eq:deld}
\delta\Delta d^{m} &= \int_0^1 dx\, x^{m} (\Delta d^{p}(x)- \Delta u^{n}(x)), \nonumber \\ 
                  &= \xm^p_{\Delta d} - \xm^n_{\Delta u}\,.
\end{align}
In the limit where the strange and light quarks have nearly equal
mass, these CSV spin moments are related to hyperon spin moments by
\begin{align}
\label{eq:csvu}
\delta\Delta u^{m} &\sim  \xm^{\Sigma}_{\Delta u} - \xm^{\Xi}_{\Delta s}\,, \\
\label{eq:csvd}
\delta\Delta d^{m} &\sim   \xm^{\Sigma}_{\Delta s} - \xm^{\Xi}_{\Delta u}\,. 
\end{align}

\section{Lattice Simulation Details}
In the numerical calculation of the moments defined in
Eqs.\eqref{eq:delu}-\eqref{eq:csvd}, our gauge field configurations
have been generated with $N_f=2+1$ flavours of dynamical fermions,
using the Symanzik improved gluon action and nonperturbatively ${\cal
  O}(a)$-improved Wilson fermions \cite{Cundy:2009yy}.
The quark masses are chosen by first finding the
SU(3)$_{\mathrm{flavour}}$-symmetric point where flavour singlet
quantities take on their physical values and then varying the
individual quark masses while keeping the singlet quark mass
$\mqbar=(m_u+m_d+m_s)/3=(2m_l+m_s)/3$
constant~\cite{Bietenholz:2010jr,Bietenholz:2011qq}.
Simulations are performed on lattice volumes of $24^3\times 48$ with
lattice spacing, $a=0.083(3)$fm.
A summary of the dynamical configurations used is given in
Table~\ref{tab:results-m}.
More details regarding the tuning of the simulation parameters can be
found in Refs.~\cite{Bietenholz:2010jr,Bietenholz:2011qq}.

On the lattice we compute moments of the spin-dependent quark
distribution functions, $\Delta q(x)$
\begin{equation}
  \label{eq:pdf}
  \langle x^{m}\rangle^B_{\Delta q} = \int^1_0 dx\,
  x^{m}(\Delta q^B(x)+(-1)^m\Delta \bar{q}^B(x))\ ,
\end{equation}
where $x$ is the Bjorken scaling variable associated with baryon $B$.
This involves calculating the matrix elements of local twist-2 operators, namely
\begin{multline}
  \label{eq:me}
  \langle B(\vec{p})|\big[{\cal
    O}_q^{5\{\mu_0\ldots\mu_m\}}-\mathrm{Tr}\big]|B(\vec{p})\rangle \\
= 2\langle x^{m}\rangle^B_{\Delta q} [s^{\{\mu_0}p^{\mu_1}\cdots
    p^{\mu_m\}} - \mathrm{Tr}],
\end{multline}
where
$  {\cal O}_q^{5\mu_0\ldots\mu_m}=i^{m}
  \bar{q}\gamma_5\gamma^{\mu_0}\Dd{\mu_1}\cdots\Dd{\mu_m}q\ .
$
We note that in the case of the unpolarised quark distribution
functions the lowest moment is protected by a sum rule (baryon number
conservation).
As a result, we only considered the first non-trivial moment, $\x_q$,
in our previous calculation of the spin-independent CSV
\cite{Horsley:2010th}.
The lowest moment of the spin-dependent quark distribution functions,
however, is not protected by such a sum rule.
Hence, in this work we consider the first two $(m=0,1)$ moments, which,
according to Eq.~(\ref{eq:pdf}), contain one C-even and one C-odd
moment.
This allows us to better assess the impact of the sea distribution in
our results.

In this paper we only consider the quark-line connected contributions
to the first two moments, $\1_{\Delta q},\,\x_{\Delta q}$, which means
that we only include the part of $\bar{q}^B$ coming from quark-line
connected backward moving quarks, the so-called Z-graphs.
While the contributions from disconnected insertions are expected to
be small \cite{Deka:2008xr,QCDSF:2011aa}, in the following analysis we
will focus on differences of baryons and so these contributions will
cancel in the SU(3)$_{\mathrm{flavour}}$ limit and should be
negligible for small expansions around this limit, as considered here.

\begin{table}[tbp]
\addtolength{\tabcolsep}{+7.0pt}
\begin{tabular}{c|c|c|c|c}
\# & $\kappa_l$ & $\kappa_s$ & $m_\pi$\,[MeV]& $m_K$\,[MeV] \\
\hline
1 & 0.12083 & 0.12104 & 460(17) & 401(15) \\
2 & 0.12090 & 0.12090 & 423(15) & 423(15) \\
3 & 0.12095 & 0.12080 & 395(14) & 438(16) \\
4 & 0.12100 & 0.12070 & 360(13) & 451(16) \\
5 & 0.12104 & 0.12062 & 334(12) & 463(17) 
\end{tabular}
\caption{Pion and kaon masses on $24^3\times 48$ lattices with lattice
  spacing, $a=0.083(3)$fm \cite{Bietenholz:2010jr}, where the error on
  the lattice spacing has been included in the errors for $m_\pi$ and
  $m_K$. The first column denotes the ensemble number.}
\label{tab:results-m}
\end{table}

We use a nucleon polarised in the $+z$-direction with the standard
local operators
\begin{equation}
{\cal O}_{\Delta q}^{\1}={\cal O}_{\Delta q}^{53} \qquad \text{and} \qquad
   {\cal O}_{\Delta q}^{\x}={\cal O}_{\Delta q}^{5\{43\}}\, .
\end{equation}
The matrix elements in Eq.~(\ref{eq:me}) are obtained on the lattice by
considering the ratios:
\begin{align}
  R(t,\tau,\vec p,{\cal O}_{\Delta q}^{\1})&= \frac{C_{\mathrm{3pt}}(t,\tau,\vec
    p,{\cal O}_{\Delta q}^{\1})}{C_{\mathrm{2pt}}(t,\vec p)} 
     =  i\1_{\Delta q}\ ,\nonumber\\
  R(t,\tau,\vec p,{\cal O}_{\Delta q}^{\x})&= \frac{C_{\mathrm{3pt}}(t,\tau,\vec
    p,{\cal O}_{\Delta q}^{\x})}{C_{\mathrm{2pt}}(t,\vec p)} 
     = -i\frac{m_B}{2} \x_{\Delta q}\, ,
  \label{eq:rat}
\end{align}
where $C_{\mathrm{2pt}}$ and $C_{\mathrm{3pt}}$ are lattice two and
three-point functions, respectively, with total momentum, $\vec p$,
(in our simulation we consider only $\vec{p}=0$).
The operators ${\cal O}_{\Delta q}^{\1}$ and ${\cal O}_{\Delta q}^{\x}$ 
are inserted into the three-point function,
$C_{\mathrm{3pt}}(t,\tau,\vec p,{\cal O})$ at time, $\tau$, between the baryon
source located at time, $t=0$, and sink at time, $t$.


\section{Lattice Results}
\begin{figure}[tbp]
\includegraphics[width=1.0\columnwidth]{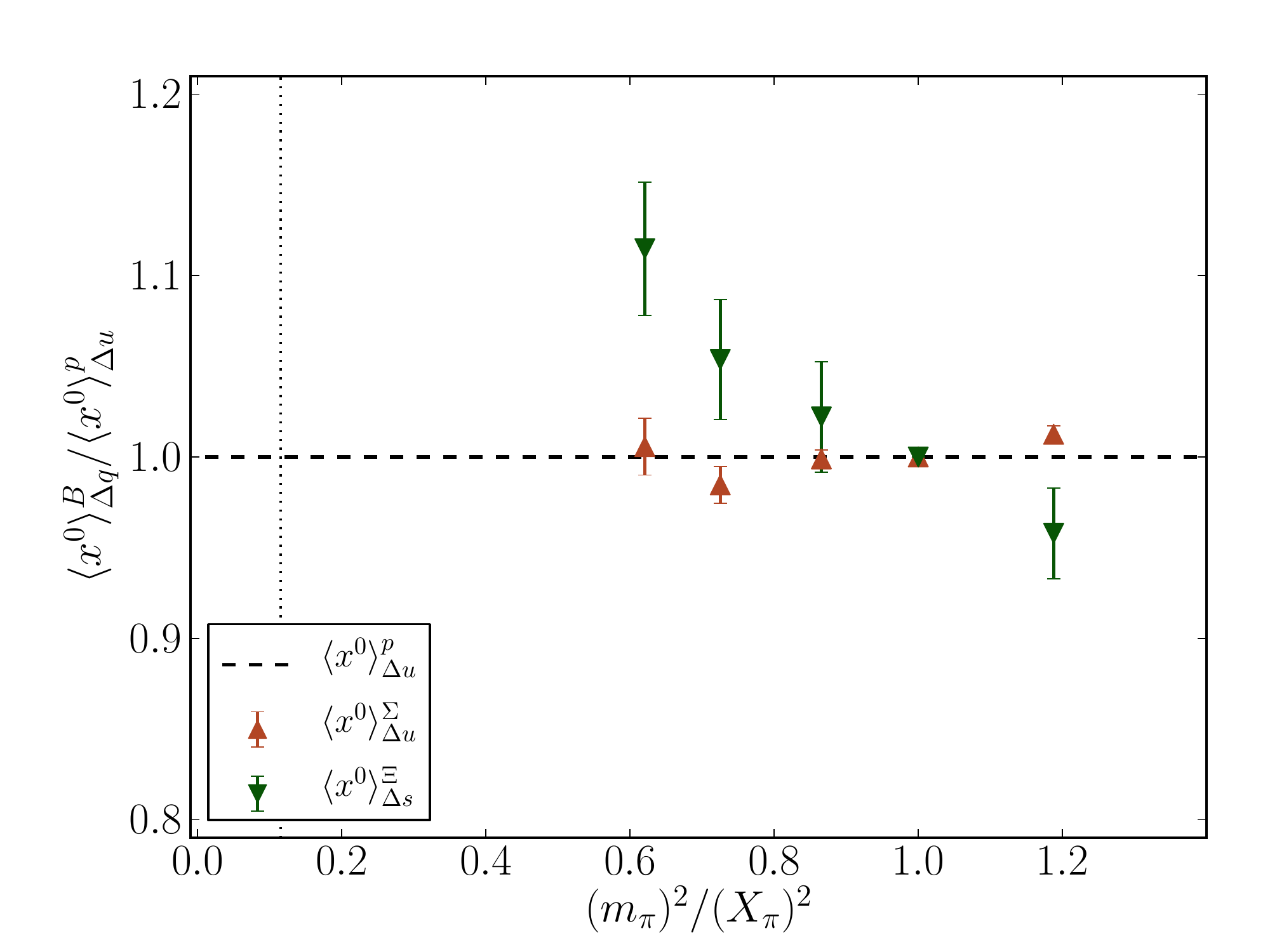}
\caption{Ratio of doubly-represented quark contributions to a baryon's
  spin, $\1_{\Delta u}^\Sigma/\1_{\Delta u}^p$ and $\1_{\Delta
    s}^\Xi/\1_{\Delta u}^p$ as a function of $m_\pi^2/X_\pi^2$, where
  the pion masses are normalised to the lattice determination of
  $X_\pi$. The vertical dotted line indicates the physical pion mass.}
\label{fig:xu-0}
\end{figure}

The operators used for determining the moments of the spin-dependent
PDFs need to be renormalised, preferably using a nonperturbative
method such as RI$^\prime$-MOM
\cite{Martinelli:1994ty,Gockeler:1998ye,Gockeler:2010yr}.
Here, however, we will only present results for ratios of the first
two moments so that the renormalisation constants cancel.
%
In Fig.~\ref{fig:xu-0} we present results for the ratio of the
$u(s)$-quark contribution to the spin of the $\Sigma(\Xi)$ baryon to
the contribution of the $u$ in the proton, as a function of $m_\pi^2$,
normalised with the centre-of-mass of the pseudoscalar meson octet,
$X_\pi=\sqrt{(2m_K^2+m_\pi^2)/3}=411$~MeV.
They are also given in Table~\ref{tab:results-1} for each ensemble.
We see that the ratio of the contribution from the $u$-quark to
the spin of the $\Sigma$ and the proton is roughly constant as the
quark mass is decreased.
We also observe that the contribution from the strange quark to the
spin of the $\Xi$-baryon is greater than that of the $u$-quark in the
proton and increases as the mass of the light (strange) quark is
decreased (increased).

\begin{figure}[tbp]
\centering\includegraphics[width=1.0\columnwidth]{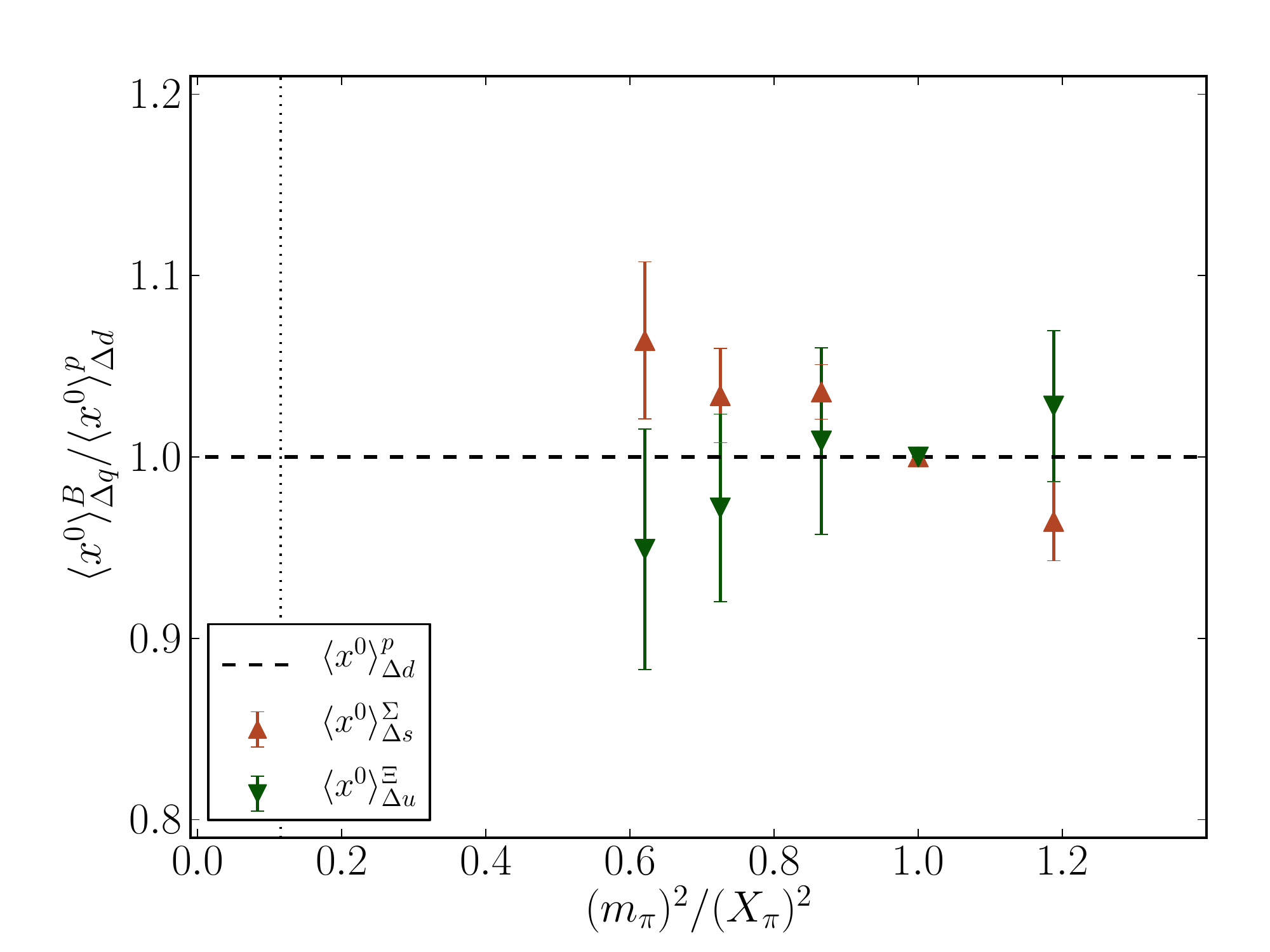}
\caption{Ratio of singly represented quark contributions to a baryon's
  spin, $\1_{\Delta s}^\Sigma/\1_{\Delta d}^p$ and $\1_{\Delta
    u}^\Xi/\1_{\Delta d}^p$ as a function of $m_\pi^2/X_\pi^2$, where
  the pion masses are normalised to the lattice determination of
  $X_\pi$. The vertical dotted line indicates the physical pion mass.}
\label{fig:xd-0}
\end{figure}

Unlike the unpolarised case in \cite{Horsley:2010th}, there is no sum
rule to preserve the total spin-dependent quark contributions.
This implies that the strange quark contribution to the spin of the
$\Sigma$ doesn't necessarily have to be the same as the $d$-quark to
the proton, and in fact we see in Fig.~\ref{fig:xd-0} that $\1_{\Delta
  s}^\Sigma>\1_{\Delta d}^p$.
Conversely, we note in Fig.~\ref{fig:xd-0} that the $u$-quark in the
$\Xi$ baryon feels the effect of the two heavier strange quarks and
$\1_{\Delta u}^\Xi<\1_{\Delta d}^p$ decreases as we approach the
physical point.

\begin{table}[tbp]
\begin{center}
\begin{tabular}{c|@{\hspace{1mm}}c@{\hspace{1mm}}|@{\hspace{1mm}}c@{\hspace{1mm}}|@{\hspace{1mm}}c@{\hspace{1mm}}|@{\hspace{1mm}}c}
\# & 
${\1_{\Delta u}^\Sigma}/{\1_{\Delta u}^p}$ &
${\1_{\Delta s}^\Sigma}/{\1_{\Delta d}^p}$ &
${\1_{\Delta s}^\Xi}/{\1_{\Delta u}^p}$ &
${\1_{\Delta u}^\Xi}/{\1_{\Delta d}^p}$ \\
\hline
1 & 1.013(5)  & 0.964(22) & 0.958(25) & 1.028(42) \\
2 & 1.0       & 1.0       & 1.0       & 1.0       \\
3 & 0.999(5)  & 1.036(15) & 1.022(30) & 1.009(51) \\
4 & 0.985(10) & 1.034(26) & 1.054(33) & 0.972(52) \\
5 & 1.006(16) & 1.064(43) & 1.115(37) & 0.949(66) 
\end{tabular}
\caption{Ratios of the zeroth moment of the C-even, spin-dependent hyperon PDFs.}
\label{tab:results-1}
\end{center}
\end{table}

\begin{figure}[tbp]
\centering\includegraphics[width=1.0\columnwidth]{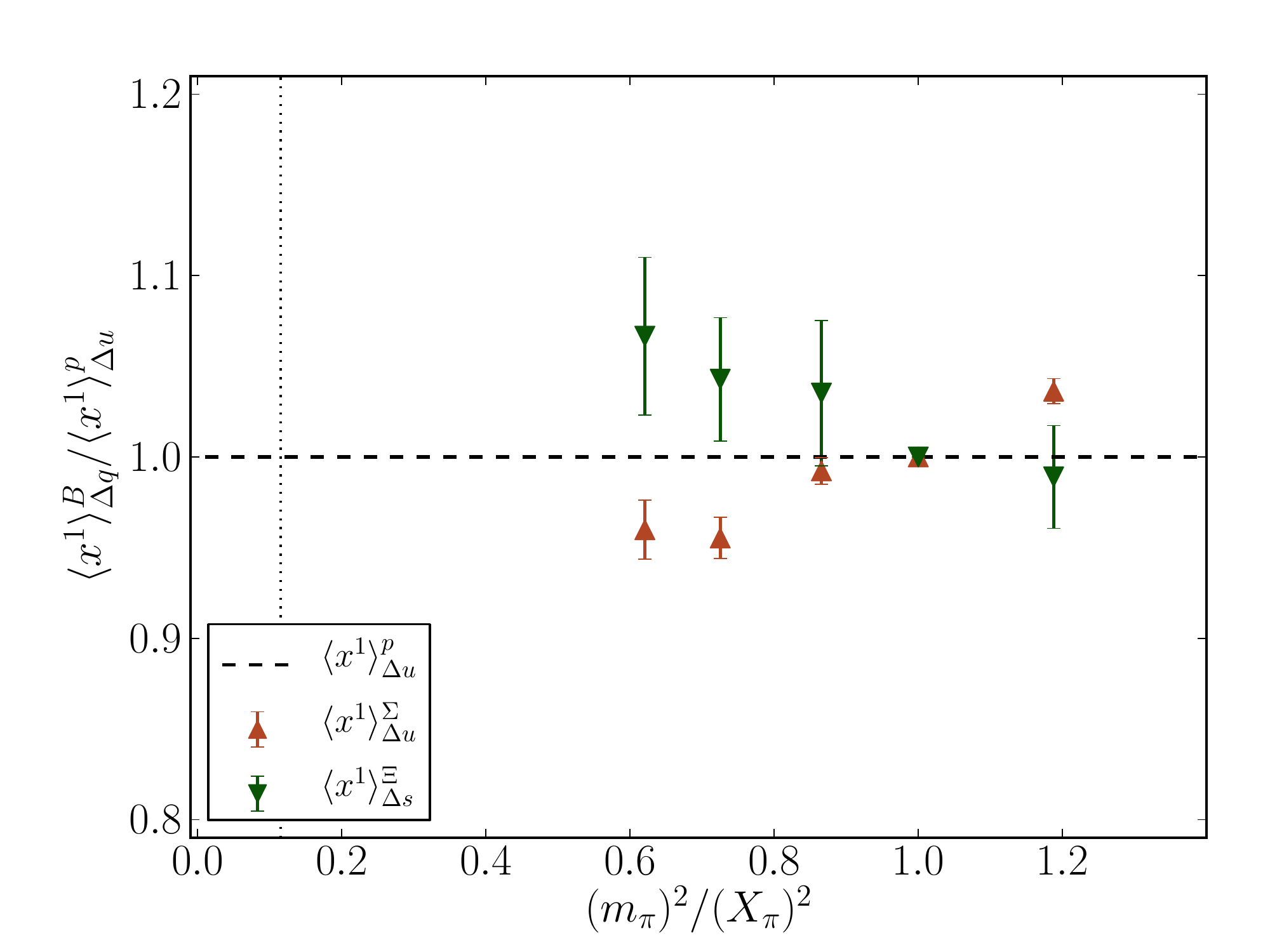}
\caption{Ratio of doubly represented spin-dependent quark momentum
  fractions, $\x_{\Delta u}^\Sigma/\x_{\Delta u}^p$ and $\x_{\Delta
    s}^\Xi/\x_{\Delta u}^p$ as a function of $m_\pi^2/X_\pi^2$, where
  the pion masses are normalised to the lattice determination of
  $X_\pi$. The vertical dotted line indicates the physical pion mass.}
\label{fig:xu-1}
\end{figure}

\begin{figure}[tbp]
\centering\includegraphics[width=1.0\columnwidth]{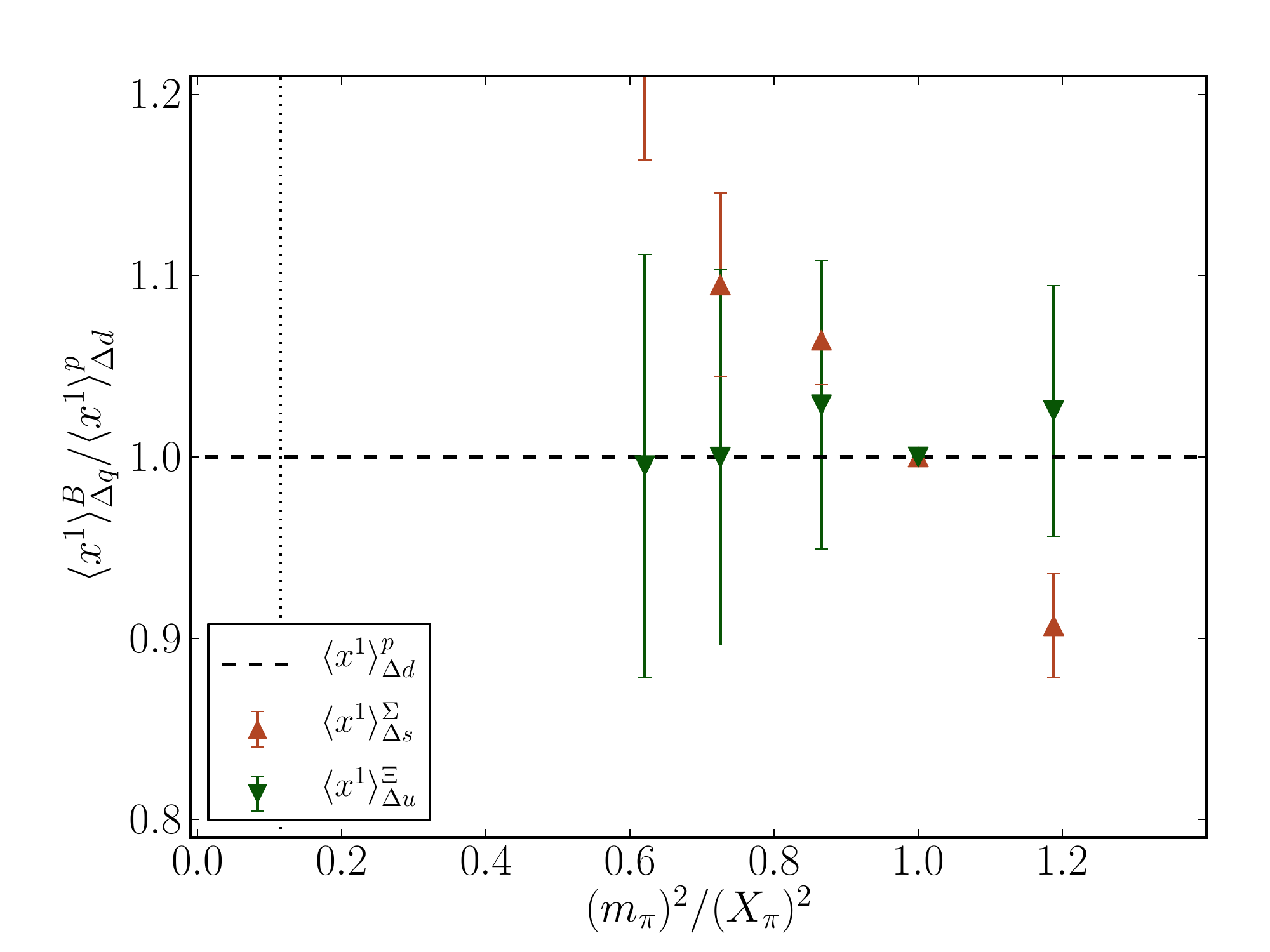}
\caption{Ratio of singly represented spin-dependent quark momentum
  fractions, $\x_{\Delta s}^\Sigma/\x_{\Delta d}^p$ and $\x_{\Delta
    u}^\Xi/\x_{\Delta d}^p$ as a function of $m_\pi^2/X_\pi^2$, where
  the pion masses are normalised to the lattice determination of
  $X_\pi$. The vertical dotted line indicates the physical pion mass.}
\label{fig:xd-1}
\end{figure}

%
Similar effects are seen in the $m=1$ (or $x$-) moments given in
Table~\ref{tab:results-x} and shown in
Fig.~\ref{fig:xu-1} and \ref{fig:xd-1}, with the exception that in
this case, $\x_{\Delta u}^\Sigma<\x_{\Delta u}^p$, i.e. the $u$-quark
in the $\Sigma$ feels the effect of the heavier strange quark being
present in the baryon.

\begin{table}[tbp]
\begin{center}
\begin{tabular}{c|@{\hspace{1mm}}c@{\hspace{1mm}}|@{\hspace{1mm}}c@{\hspace{1mm}}|@{\hspace{1mm}}c@{\hspace{1mm}}|@{\hspace{1mm}}c}
\# & 
${\x_{\Delta u}^\Sigma}/{\x_{\Delta u}^p}$ &
${\x_{\Delta s}^\Sigma}/{\x_{\Delta d}^p}$ &
${\x_{\Delta s}^\Xi}/{\x_{\Delta u}^p}$ &
${\x_{\Delta u}^\Xi}/{\x_{\Delta d}^p}$ \\
\hline
1 & 1.036(7)  & 0.907(29) & 0.989(28) & 1.025(69) \\
2 & 1.0       & 1.0       & 1.0       & 1.0       \\
3 & 0.992(7)  & 1.064(24) & 1.035(40) & 1.029(79) \\
4 & 0.956(11) & 1.095(51) & 1.043(34) & 1.000(104)\\
5 & 0.960(16) & 1.257(93) & 1.067(43) & 0.995(117)
\end{tabular}
\caption{Ratios of the first moment of the C-odd spin-dependent hyperon
  PDFs.}
\label{tab:results-x}
\end{center}
\end{table}

To infer the level of CSV relevant to the nucleon, we only need to
consider a small expansion about the
SU(3)$_{\mathrm{flavour}}$ symmetric point, for which linear flavour
expansions prove to work extremely well \cite{Bietenholz:2010jr}.
For instance, we can write
\begin{align}
\delta\Delta u^0 &= m_\delta\left(-\frac{\del\1_{\Delta u}^p}{\del m_u}
+\frac{\del\1_{\Delta u}^p}{\del m_d}\right)+\cO(m_\delta^2)\,,\nonumber\\
\delta\Delta u^1 &= m_\delta\left(-\frac{\del\x_{\Delta u}^p}{\del m_u}
+\frac{\del\x_{\Delta u}^p}{\del m_d}\right)+\cO(m_\delta^2)\,,
\label{eq:du}
\end{align}
where $m_\delta\equiv (m_d-m_u)$ and we have already made use of
charge symmetry by equating 
$\del\xm_{\Delta d}^n/\del m_d=\del\langle
x^m\rangle_{\Delta u}^p/\del m_u$ and 
$\del\xm_{\Delta d}^n/\del m_u=\del\langle
x^m\rangle_{\Delta u}^p/\del m_d$.
Similar expressions hold for $\delta\Delta d^0$ and $\delta\Delta d^1$.

Near the SU(3)$_{\mathrm{flavour}}$ symmetric point, we note that the
up quark in the proton is equivalent to an up quark in a $\Sigma^+$ or
a strange quark in a $\Xi^0$, which we describe collectively as the
``doubly-represented'' quark \cite{Leinweber:1995ie}.

The local derivatives required for $\delta\Delta u^m$ can be obtained by
varying the masses of the up and down quarks independently. 
Within the present calculation, we note that the difference
$\xm_{\Delta s}^{\Xi}-\xm_{\Delta u}^p$ measures precisely the
variation of the doubly-represented quark matrix element with respect
to the doubly-represented quark mass (while holding the
singly-represented quark mass fixed). 
Similar variations allow us to evaluate the other required
derivatives, where we write
\begin{align}
\frac{\del \xm_{\Delta u}^p}{\del m_u}&\simeq
\frac{\xm_{\Delta s}^{\Xi^0}-\xm_{\Delta u}^p}{m_s-m_l}\,,\nonumber\\
\frac{\del \xm_{\Delta u}^p}{\del m_d}&\simeq 
\frac{\xm_{\Delta u}^{\Sigma^+}-\xm_{\Delta u}^p}{m_s-m_l}\,,\nonumber\\
\frac{\del \xm_{\Delta d}^p}{\del m_u}&\simeq
\frac{\xm_{\Delta u}^{\Xi^0}-\xm_{\Delta d}^p}{m_s-m_l}\,,\nonumber\\ 
\frac{\del \xm_{\Delta d}^p}{\del m_d}&\simeq
\frac{\xm_{\Delta s}^{\Sigma^+}-\xm_{\Delta d}^p}{m_s-m_l}\,.
\end{align}
With these expressions and Eq.~(\ref{eq:du}), we obtain the relevant
combinations for our determination of CSV in the nucleon
\begin{align}
\delta\Delta u^m &= m_\delta\frac{\xm_{\Delta u}^{\Sigma^+}-\xm_{\Delta s}^{\Xi^0}}{m_s-m_l}\,,\ \nonumber\\
\delta\Delta d^m &= m_\delta\frac{\xm_{\Delta s}^{\Sigma^+}-\xm_{\Delta u}^{\Xi^0}}{m_s-m_l}\,.
\end{align}

\begin{figure}[tbp]
\centering\includegraphics[width=1.0\columnwidth]{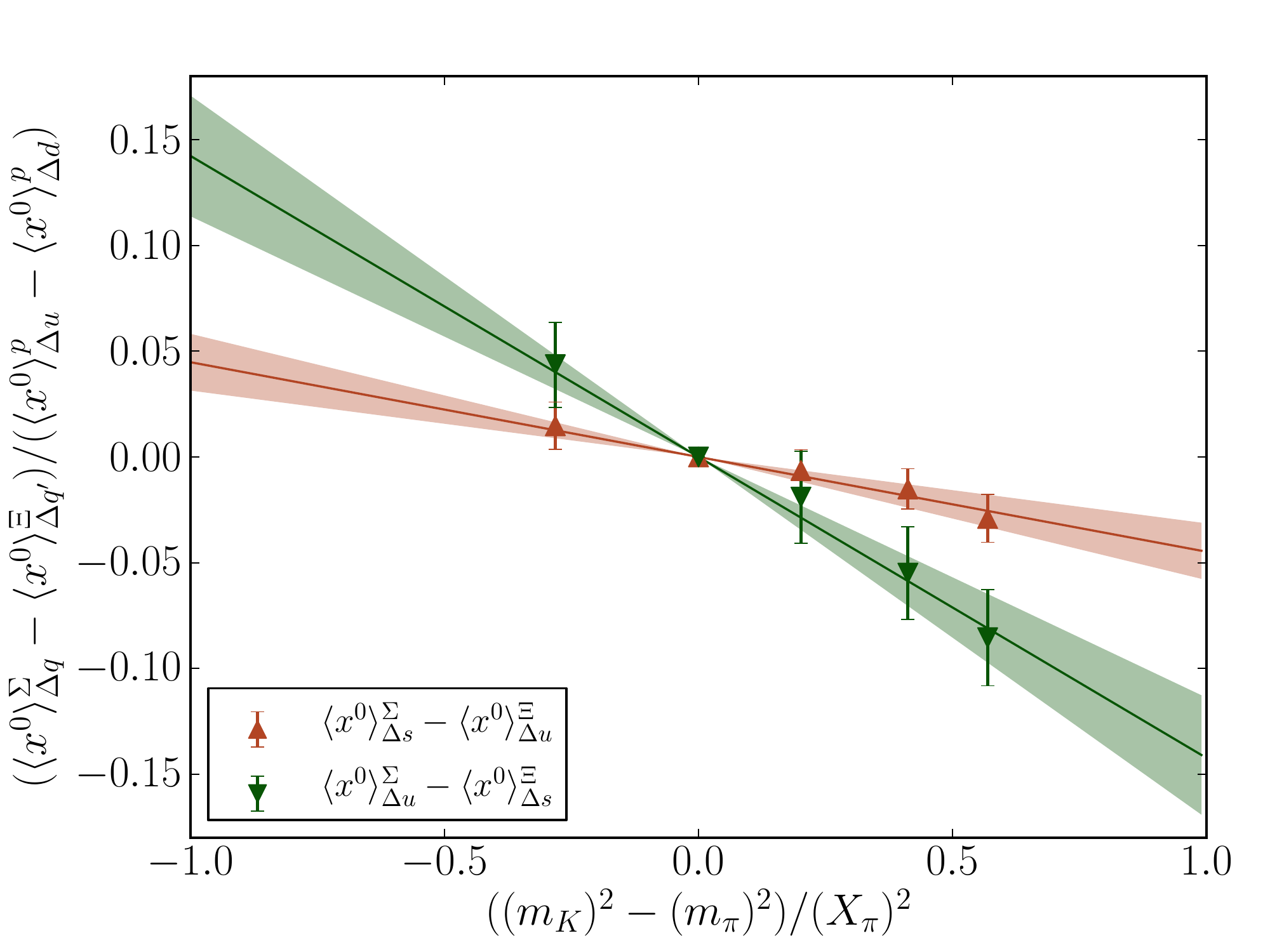}
\caption{The zeroth moment of the spin difference between doubly and singly represented
  quarks in the $\Sigma$ and $\Xi$ as a function of the strange/light
  quark mass difference. We deduce $\delta\Delta u^0$ and $\delta\Delta d^0$,
  respectively, from the slopes of these curves
  (c.f. Eqs.~(\ref{eq:deltau0}) - (\ref{eq:deltad1})).}
\label{fig:spin0}
\end{figure}

\begin{figure}[tbp]
\centering\includegraphics[width=1.0\columnwidth]{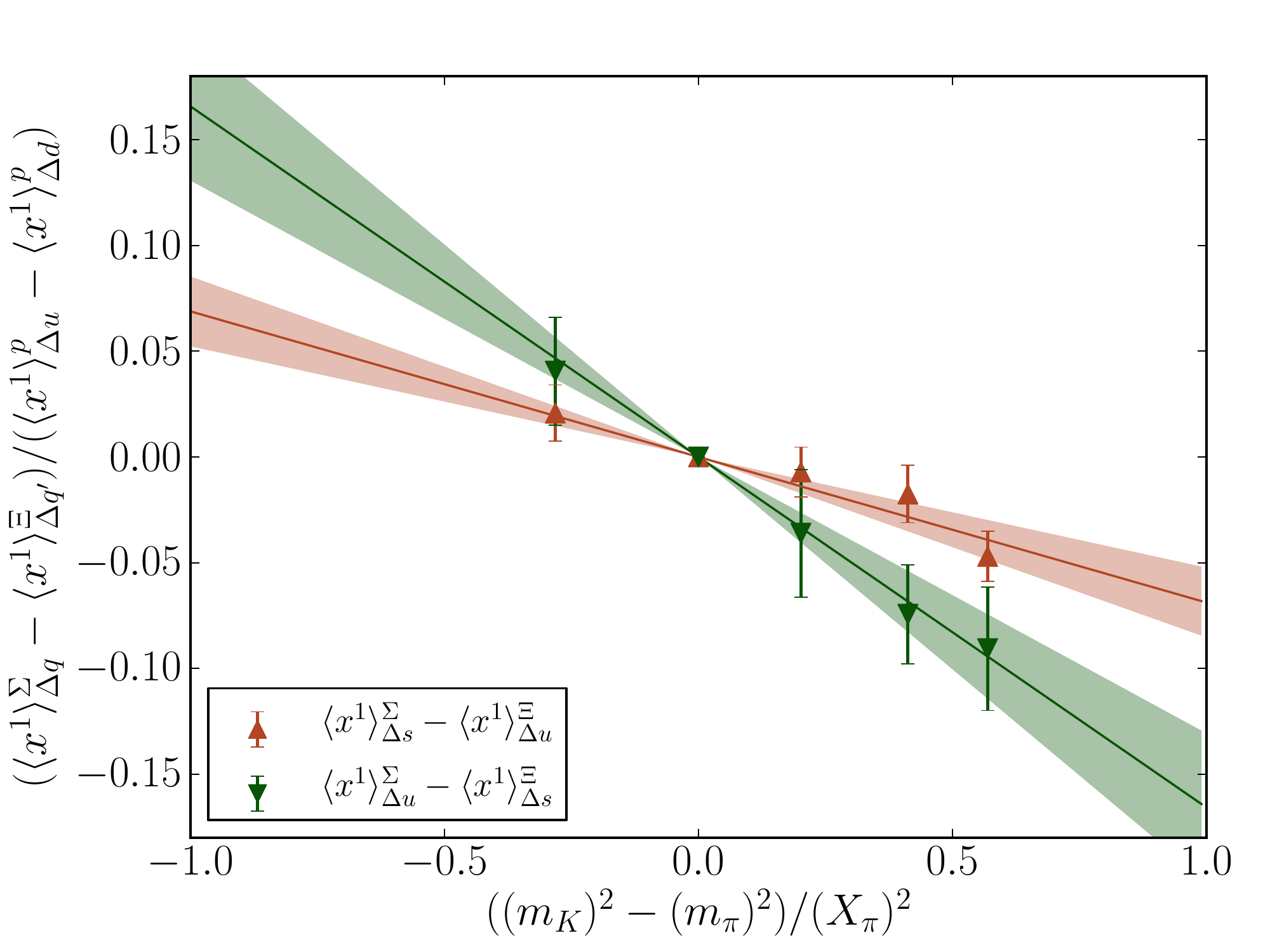}
\caption{The first moment of the spin difference in the $\Sigma$ and $\Xi$ vs. 
the strange/light quark mass difference, 
from which we deduce $\delta\Delta u^1$ and $\delta\Delta d^1$.}
\label{fig:spin1}
\end{figure}

By invoking the Gell-Mann--Oakes--Renner relation and normalising to
the total nucleon isovector $m=0,1$ spin-dependent moments, we write
\begin{align}
\frac{\delta\Delta u^m}{\xm_{\Delta u-\Delta d}^p} &= \frac{m_\delta}{\mqbar}
\frac{(\xm_{\Delta u}^{\Sigma^+}-\xm_{\Delta s}^{\Xi^0})/\xm_{\Delta u-\Delta d}^p}{(m_K^2-m_\pi^2)/X_{\pi}^2}\,,\\
\frac{\delta\Delta d^m}{\xm_{\Delta u-\Delta d}^p} &= \frac{m_\delta}{\mqbar}
\frac{(\xm_{\Delta s}^{\Sigma^+}-\xm_{\Delta u}^{\Xi^0})/\xm_{\Delta u-\Delta d}^p}{(m_K^2-m_\pi^2)/X_{\pi}^2}\,.
\end{align}
Written in this way, the fractional spin-dependent CSV terms are just
the slopes of the curves shown in Figs.~\ref{fig:spin0} and
\ref{fig:spin1} (evaluated at the symmetry point) multiplied by the
ratio $m_\delta/\mqbar$. By fitting the slopes, we obtain
\begin{align}
\frac{\delta\Delta u^{0+}}{\1_{\Delta u^+-\Delta d^+}^p}&= \frac{m_\delta}{\mqbar}(-0.137\pm 0.028)\,,
\label{eq:deltau0} \\
\frac{\delta\Delta d^{0+}}{\1_{\Delta u^+-\Delta d^+}^p}&= \frac{m_\delta}{\mqbar}(-0.0433\pm 0.013)\,,
\label{eq:deltad0} \\
\frac{\delta\Delta u^{1-}}{\x_{\Delta u^--\Delta d^-}^p}&= \frac{m_\delta}{\mqbar}(-0.161\pm 0.035)\,,
\label{eq:deltau1} \\
\frac{\delta\Delta d^{1-}}{\x_{\Delta u^--\Delta d^-}^p}&= \frac{m_\delta}{\mqbar}(-0.068\pm 0.016)\,.
\label{eq:deltad1}
\end{align}

Chiral perturbation theory yields the quark mass ratio
$m_\delta/\mqbar=0.066(7)$ \cite{Leutwyler:1996qg}
while the experimentally determined moments are $\1_{\Delta u^+-\Delta
  d^+}=g_A=1.2695(29)$ \cite{Nakamura:2010zzi} and $\x_{\Delta u^-
  -\Delta d^-}^p= 0.190(8)$ \cite{Blumlein:2010rn} in the \msbar
scheme at $4\gev^2$.
We note that in principle the zeroth moments in
Eqs.~(\ref{eq:deltau0}) and (\ref{eq:deltad0}) will receive their scale
dependence from an addtional term 
\be
\frac{z(\mu,a)}{3}\frac{(\delta\Delta u^{0} + \delta\Delta
  d^{0})}{\1_{\Delta u-\Delta d}^p}\ ,
\ee
where $z(\mu,a)$ is the difference between the (scale-dependent)
singlet and (scale-independent) nonsinglet axial-vector current
renormalisation constants.
At order ${\cal O}(\alpha_s^2)$ in perturbation theory this results in
a correction of $<1\%$ at $\mu^2=a^{-2}=6\,\mathrm{GeV}^2$
\cite{Skouroupathis:2008mf,QCDSF:2011aa} and, due to its small
anomalous dimension, also at other scales,
e.g. $\mu^2=4\,\mathrm{GeV}^2$.

Substituting these values into
Eqs.~(\ref{eq:deltau0})-(\ref{eq:deltad1}) yields the first lattice
QCD estimates of the spin CSV moments
\begin{align}
\delta\Delta u^{0+} &= -0.0116 (27), & \delta\Delta d^{0+} &= -0.0036 (11) \, , \nonumber \\  
\delta\Delta u^{1-} &= -0.0020 (5),  & \delta\Delta d^{1-} &= -0.0009 (2) \,.
\label{eq:DelmomLat}
\end{align}
We can make several observations regarding these spin CSV
moments. 
First, the fractional spin CSV for both moments and both flavours are
similar in magnitude and all have the same (negative) sign.
Second, we can compare the first moments of the spin CSV distributions
with the corresponding first moments of the spin-independent CSV
distributions that were reported in Ref.~\cite{Horsley:2010th}, namely
\be \delta u^+ =
-0.0023(6), \quad \delta d^+ = +0.0020(3) \,.
\label{eq:latmom}
\ee 
The first moments of the spin-independent CSV results have roughly
equal magnitudes but opposite signs, with $\delta u$ being negative
and $\delta d$ positive, in both qualitative and quantitative
agreement with quark model
predictions~\cite{Rodionov:1994cg,Londergan:2003pq} and with the
best-fit values from a global fit that included valence
CSV~\cite{Martin:2003sk}.

Next we note that the zeroth moments of the spin-dependent CSV
distributions are larger than the first moments.
Lastly, we have estimated the CSV associated only with the $u-d$ mass
difference.
It is important to also find a method to investigate the CSV induced
by electromagnetic effects which, at least in the unpolarised case, is
expected to be of a similar size \cite{Martin:2004dh,Gluck:2005xh}.

\section{Quark Model Computation}
We can compare the lattice results with estimates of valence quark
spin-dependent PDFs obtained from quark model calculations. Schreiber, Signal
and Thomas \cite{Schreiber:1991tc} calculated parton spin
distributions from bag models. 
Sather \cite{Sather:1991je} derived an analytic approximation giving
valence parton CSV distributions in terms of derivatives of
phenomenological PDFs.
Sather's equations are valid for parton distributions at a low $Q^2$
scale appropriate for quark models, and should also be valid for CSV
spin distributions. 
In this approximation, the valence parton CSV spin distributions are
\begin{align}
\delta\Delta d^-(x) &= -\frac{\delta M}{M}\,\frac{d}{dx}\left[x\Delta d^-(x)\right] 
     -\frac{\delta m}{M}\,\frac{d}{dx}\,\Delta d^-(x), \nonumber \\ 
\delta\Delta u^-(x) &= \frac{\delta M}{M}\,
  \left(-\Delta u^-(x) + (1-x)\frac{d}{dx}\,\Delta u^-(x)\right),
  \label{eq:Satspin}  
\end{align}
where $\delta M$ is the $n-p$ mass difference and $\delta m$ is the
diquark mass difference $m_{dd} - m_{uu}$ which is determined rather
accurately to be 4 MeV \cite{Bickerstaff:1989ch}.
The zeroth moment of the spin dependent CSV distributions is overly
sensitive to the small-$x$ behaviour of these PDFs, a region where the
quark model results are less reliable.
Therefore we compare only with the first moments of the CSV spin-dependent
distributions. Using the model of Schreiber, Signal and Thomas we find
\be
\delta\Delta u^{1-} = -0.0008 , \quad \delta\Delta d^{1-} = -0.0011.
\label{eq:Delmom}
\ee
Alternatively, if we use the spin-dependent PDFs from a
Nambu-Jona-Lasinio model calculation \cite{Cloet:2005pp,Cloet:2005rt}
together with Eqs.~(\ref{eq:Satspin}), we find
\be
\delta \Delta u^{1-} =  -0.0003\,, \quad
\delta \Delta d^{1-} =  -0.0007\,.
\ee
These phenomenological model predictions agree with the lattice
results insofar as the first spin moments are all the same (negative)
sign, and have similar magnitudes.
As for the spin independent case, the result for the down
distribution, which is dominated by the diquark mass shift, is in
better agreement than that for the up quark where there are a number
of small corrections, not all included in the Sather approximation.

\section{Bjorken Sum Rule}
Finally, the spin-dependent CSV distributions contribute to the Bjorken sum
rule \cite{Bjorken:1966}, which has the form
\begin{multline}
\int_0^1 dx \left[ g_1^p(x) - g_1^n(x) \right] = \frac{G_A}{6G_V} \, [1- \frac{\alpha_S(Q^2)}{\pi}] \\
= \langle \, \frac{\Delta u^+(x) - \Delta d^+(x)}{6} +
\frac{4\delta\Delta d^+(x) + \delta\Delta u^+(x)}{18}
\,\rangle.
\label{eq:Bjsum}
\end{multline}
In the first line of Eq.~(\ref{eq:Bjsum}) we write the Bjorken sum
rule in terms of the difference of the spin-dependent structure functions $g_1$ for
the proton and neutron, integrated over all $x$. 
In the second line of Eq.~(\ref{eq:Bjsum}) we write the sum rule in
terms of the first moment of spin-dependent parton distributions.
This quantity is correct up to terms of order ${\cal O}(\alpha_S)$
(there are also higher-twist terms of order ${\cal O}(1/Q^2)$). 
We see that, except for the CSV corrections, this ratio is given by
the zeroth moment of the difference of the C-even spin distributions
$\Delta u^+$ and $\Delta d^+$ integrated over all $x$.

We have included the contribution from partonic spin CSV in
Eq.~(\ref{eq:Bjsum}) which is noteworthy for several reasons.
First, with the exception of corrections arising from partonic spin
CSV terms, there are essentially no other partonic corrections to the
Bjorken sum rule at leading twist (this is one reason why it is so
important to obtain precise values for this sum rule).
Second, the correction involves the zeroth moments of $\delta\Delta
u^+$ and $\delta\Delta d^+$. 
At present the Bjorken sum rule is best determined from a recent
COMPASS experiment at $Q^2 = 3$ GeV$^2$ to a precision of about 8\%
\cite{Alekseev:2010hc}.  
Using the zeroth moment obtained from our lattice calculations (see
Eq.~(\ref{eq:DelmomLat})) we estimate that the spin CSV terms
contribute approximately 1\% to the Bjorken sum rule. 
At the present measured precision it is not possible to observe such a
small contribution.
However, the Bjorken sum rule could in principle be measured at a
future electron collider, where one could imagine aiming for 1\%
precision \cite{Sichtermann:2007}.  
With such precision it is possible that the spin CSV contributions
calculated here would be sufficiently large to make a measurable
difference in the sum rule.

\section{Conclusion}
In summary, we have performed the first lattice determinations of the
polarised quark moments of the hyperons, $\Sigma$ and $\Xi$ in
$N_f=2+1$ lattice QCD.
By examining the SU(3)$_{\mathrm{flavour}}$-breaking effects in these
momentum fractions, we are able to extract the first QCD determination
of the size and sign of charge-symmetry violation in the
spin-dependent parton distribution functions in the nucleon,
$\delta\Delta u$ and $\delta\Delta d$.
We compare our results with estimates of the first moment of the
parton spin CSV from a quark model calculation, obtaining qualitative
agreement with the quark model results.
Finally, we estimate the contribution of partonic spin CSV to the
Bjorken sum rule, and show that spin CSV effects should change the
Bjorken sum rule by approximately 1\%.

\section*{Acknowledgements}
The numerical calculations have been performed on the apeNEXT at
NIC / DESY (Zeuthen, Germany), the IBM BlueGeneL at EPCC (Edinburgh,
UK), the BlueGeneP (JuGene) and the Nehalem Cluster (JuRoPa) at NIC
(J\"ulich, Germany), and the SGI Altix and ICE 8200 systems at LRZ
(Munich, Germany) and HLRN (Berlin-Hannover, Germany).
We have made use of the Chroma software suite \cite{Edwards:2004sx}.
The BlueGene codes were optimised with Bagel \cite{Bagel}.
This work has been supported in part by the DFG (SFB/TR 55, Hadron
Physics from Lattice QCD) and the EU under grants 238353
(ITN STRONGnet) and 227431 (HadronPhysics2).
JTL is supported by the US National Science Foundation grant NSF PHY-0854805. 
This work was also supported by the University of Adelaide and the Australian
Research Council through an Australian Laureate Fellowship (FL0992247,
AWT), a Future Fellowship (FT100100005, JMZ) and Discovery Grant
(DP110101265, RDY).
%
%



\end{document}